\documentstyle[twocolumn,prc,aps,epsf,floats]{revtex}

\newcommand{\carbon}{$^{12}$C}
\newcommand{\boron}{$^{11}$B}
\newcommand{\eep}{$(e,e'p)$}
\newcommand{\mevc}{MeV/$c$}
\newcommand{\gevcsq}{(GeV/$c)^2$}
\newcommand{\emiss}{$E_m$}
\newcommand{\pmiss}{$p_m$}
\newcommand{\onep}{$1p$}
\newcommand{\ones}{$1s$}
\newcommand{\QSQ}{$Q^2$}
\newcommand{\qom}{($q,\omega$)}
\newcommand{\rhopm}{$\sigma_{red}({\bf p}_{ m},{\bf p} ')$}
\newcommand{\sigcc}{$\sigma_{ep}^{cc1}$}
\newcommand{\signr}{$\sigma_{ep}^{NR}$}
\newcommand{\sonep}{$S_{1p}$}
\newcommand{\sones}{$S_{1s}$}
\newcommand{\tp}{$T_p$}
\newcommand{\onepthree}{$1p_{3/2}$}
\newcommand{\onepone}{$1p_{1/2}$}
\newcommand{\trans}{$T$}
\newcommand{\transp}{$T_{1p}$}
\newcommand{\transs}{$T_{1s}$}
\newcommand{\etal}{{\it et al.}}
\newcommand{\emup}{$E_m^{up}$}

\begin{document}

\title{The Transparency of $^{12}$C for Protons}

\author{L. Lapik\'{a}s,$^1$ G. van der Steenhoven,$^1$
L. Frankfurt,$^2$ M.~Strikman,$^{3,4}$ \thanks{on leave of absence from PSU}
and M. Zhalov$^5$}

\address{
$^1$ Nationaal Instituut voor Kernfysica en Hoge-Energiefysica (NIKHEF),\\
P.O.~Box~41882, 1009~DB~Amsterdam, The~Netherlands\\
$^2$ School of Physics and Astronomy, Tel Aviv University, 69978 
Tel Aviv, Israel\\
$^3$ Department of Physics, Pennsylvania State University, 
University Park, PA 16802, USA\\
$^4$ Deutsches Elektronen Synchrotron DESY, Germany\\
$^5$ Institute for Nuclear Physics, St. Petersburg, Russian Federation
}

\maketitle

\begin{abstract}
Existing cross-section data for \onep -shell knockout in the reaction 
\carbon\eep\boron\ - as obtained under different kinematic conditions 
- are shown to be mutually consistent, apart from a recent measurement 
performed in Mainz.  New data have been collected at the Amsterdam 
Pulse Stretcher that confirm the normalization of the older 
measurements.  An analysis of the world's \carbon\eep\boron\ data has 
yielded precise values of the spectroscopic factor for \onep -shell 
and \ones -shell knockout from \carbon.  These values have been used 
to evaluate the transparency of the \carbon\ nucleus for \onep -shell 
and \ones -shell protons separately on the basis of recent high-energy 
\carbon\eep\boron\ data taken at a four-momentum transfer squared 
\QSQ\ of 1.1 \gevcsq.  As the resulting average value of the nuclear 
transparency, 0.81 $\pm$ 0.04, is considerably higher than the value 
obtained from previous analyses and theoretical estimates, the high 
\QSQ\ data were used instead for an independent determination of the 
spectroscopic strength for \onep\ + \ones\ knockout.  Combining these 
results with the low \QSQ\ data the spectroscopic factors appear to be 
momentum-transfer dependent.  Possible explanations of these 
surprising results in terms of reaction-mechanism effects or a 
possible breakdown of the quasi-particle concept at high \QSQ\ are 
discussed as well.

\vspace{0.4 cm}
\noindent{PACS number(s): 21.10.Jx, 21.30.Fe, 24.10.Ht, 27.20.+n}
\vspace{0.4 cm}

          
\end{abstract}

\begin{center}

\end{center}

\section{Introduction}

Electron-induced proton knockout experiments in the quasi-elastic
domain are commonly used to study single-particle properties of nuclei
\cite{lap93,stew91}.  The data set for such \eep\ measurements on
\carbon\ in particular is quite sizable
\cite{ama67,nak76,mou76,ber82,ste88a,ste88b}, possibly because the
energy-resolution requirements are modest ($ \leq $ 1 MeV) and the
target handling is easy.  Therefore, it is no surprise that
\carbon\eep\ measurements are often among the first calibration
experiments to be carried out at new high-duty factor electron
accelerators in the intermediate energy domain, such as AmPS
\cite{wit93}, TJNAF \cite{gees} and Mainz \cite{blo95}.

An early comparison of part of the world's \carbon\eep\ data for
knockout from the \onep -shell in the quasi-elastic domain
\cite{ste89} demonstrated the mutual consistency of these data.  On
the other hand, recent \carbon\eep\ data collected in Mainz
\cite{blo95} suggest that the normalization of previous data was off
by 22\%.  It is important to resolve this discrepancy for the
following reasons.  First, the spectroscopic factors derived from
\eep\ data on \carbon\ (and other nuclei) were shown to be quenched by
about 30-40\% as compared to mean-field values
\cite{lap93,stew91}, which has been interpreted as evidence for the
existence of strong correlations between nucleons in nuclei
\cite{dic92,mah91}.  A further reduction of the spectroscopic factors
by 22\% would make the commonly accepted many-body interpretation
uncertain.  Secondly, the spectroscopic factors for \onep\ and \ones
-knockout from \carbon\ enter directly into the determination of the
nuclear transparency, as recently studied on \carbon\ (and several
other nuclei) in the (squared) four-momentum transfer (\QSQ) range 1-6
\gevcsq\ in a search for color transparency phenomena
\cite{gees,mak94}.

For these reasons we have re-analyzed all existing \carbon\eep\ data 
for knockout from the \onep- and \ones -shell that were taken in the 
quasi-elastic domain at \QSQ\ $<$ 0.4 \gevcsq\ in one consistent 
approach.  The results of this analysis indicate that the 
normalization of the Mainz data set \cite{blo95} deviates with respect 
to all other existing data.  In order to further corroborate this 
finding three new \carbon\eep\ measurements were performed at the AmPS 
facility of NIKHEF in kinematics that were chosen, as close as 
possible, to resemble the kinematics used in Refs.  \cite{ste88a} and 
\cite{blo95}.  The new data are not in agreement with the Mainz 
results, but are in good agreement with all other data sets available.

Having thus established a reliable value of the spectroscopic factors 
for \onep\ and \ones\ knockout from \carbon, we reconsider the 
determination of the transparency of \carbon\ for protons.  The 
relatively large transparency values derived from this analysis 
possibly indicate that the spectroscopic factors obtained at low \QSQ\ 
can not be applied for the interpretation of high \QSQ\ measurements.  
Instead, we have used the high \QSQ\ data to study the \QSQ\ 
dependence of the total spectroscopic strength for \onep\ + \ones\ 
knockout from 0.1 to 10 \gevcsq.  An unexpected momentum-transfer 
dependence of the spectroscopic strength is observed.  We discuss 
reaction-mechanism effects and a possible breakdown of the 
quasi-particle concept at high \QSQ\ as possible explanations for this 
observation.

This paper is organized as follows: in section \ref{sec:datasets} 
details are presented of the data sets used in the analysis.  In 
section \ref{sec:onep} we describe the analysis of \onep-knockout data 
and present the new \carbon\eep\ measurements performed at AmPS. The 
analysis of \ones-knockout data is described in section 
\ref{sec:ones}.  In section \ref{sec:slacdata} a re-evaluation is 
presented of the nuclear transparency derived from the experimental 
search for color-transparency effects at SLAC (experiment NE18 
\cite{mak94}) using the magnitude of the \onep\ and \ones\ 
spectroscopic factors for the reaction \carbon\eep\ as derived in 
sections \ref{sec:onep} and \ref{sec:ones}.  The alternative 
interpretation of these data in terms of a possible \QSQ\ dependence 
of the spectroscopic strength in \carbon\ is presented in section 
\ref{sec:qdependence}, while some possible explanations for the 
observed \QSQ\ dependence are discussed in sections 
\ref{sec:qdependence} and \ref{sec:discussion}.  A summary is 
presented in section \ref{sec:summary}.

\section{Data sets}
\label{sec:datasets}

Experimental data for the cross section of the reaction \carbon\eep\ 
were obtained at Frascati \cite{ama67}, Tokyo \cite{nak76}, Saclay 
\cite{mou76,ber82}, NIKHEF 
\cite{ste88a,ste88b,ste86,bob92,ste93,kes95,kes96}, MIT/Bates 
\cite{lou86,ulm87,wein,bag89,mor99}, Mainz \cite{blo95}, SLAC 
\cite{mak94} and TJNAF \cite{gees}.  In the re-analysis of these data 
we have only used data sets covering a large ($>$ 100 \mevc) range of 
missing momentum, as this gives a good indication of the internal 
consistency of each data set.  Also, we require that the results of 
the data analysis be presented in terms of absolute cross sections 
(thus excluding Ref.  \cite{ama67}), and be centered at the low and 
intermediate missing-momentum range, i.e.  $|{\bf p}_{ m} | <$ 300 
\mevc, where most of the cross section resides.  The characteristics 
of the remaining data sets are summarized in Table \ref{tab:kinem}.

\begin{table}
\caption{Kinematics of \carbon\eep\ data sets for \onep- and
\ones-knockout discussed in the present paper.  The columns represent
data set, (range of) incident electron energy, range in missing
energy, range in missing momentum, kinetic energy of the emitted
proton, type of kinematics (parallel or \qom\-constant) and
four-momentum transfer squared.}
\begin{center}
\begin{tabular}{|l|c|c|c|c|c|c|}\hline
data set& $E_0$ & $\Delta E_m$ & $\Delta p_m$ & \tp & Kine- & \QSQ \\
       & MeV & MeV      & \mevc & MeV & atics & \gevcsq \\
\hline
\multicolumn{7}{|c|}{\onep -knockout}\\
\hline
Tokyo \cite{nak76} & 700 & 6-30 & 0,230 & 159 & \qom & 0.29\\
Saclay \cite{mou76} & 497 & 15-22 & 0,310 & 87 & \qom & 0.16\\
Saclay \cite{ber82} & 500 & 15-22 & -145,155 & 99 & par. & 0.09-0.32\\
Saclay \cite{ber82} & 500 & 15-22 & -155,165 & 99 & \qom & 0.09-0.32\\
NIKHEF \cite{ste88a}& 285-481 & g.s & -175,230 & 70 & par. &
0.02-0.26\\
Mainz \cite{blo95} & 855 & g.s. & 110,190 & 93 & par. & 0.08-0.28 \\
                   & 855 & g.s. & 70,140 & 85 & par. & 0.08-0.28\\
SLAC \cite{mak94} & 2015 & 6-25 & -180,290 & 600 & \qom & 1.11\\
\hline
\multicolumn{7}{|c|}{\ones -knockout}\\
\hline
Tokyo \cite{nak76} & 700 & 21-66 & 0,230 & 136 & \qom & 0.29 \\
Saclay \cite{mou76} & 497 & 30-50 & 0,310 & 87 & \qom & 0.16 \\
NIKHEF \cite{ste88a}& 285-481 & 30-39 & -175,230 & 70 & par. &
0.02-0.26 \\
SLAC \cite{mak94} & 2015 & 30-80 & -180,290 & 600 & \qom & 1.11\\
\hline
\end{tabular}
\end{center}
\label{tab:kinem}
\end{table}

The existing data are compared on the level of the reduced cross
section \rhopm, which is obtained from the \eep\ cross section by
dividing out the off-shell electron-proton cross section (and a
kinematic factor) and integrating the resulting spectrum
over the width of the energy intervals considered.  In many
analyses the off-shell $e-p$ cross section \sigcc\ of Ref.
\cite{for83} has been used, whereas in Refs.  \cite{nak76,mou76,ber82},
for instance, a different prescription \cite{mougth,potter} is used.
Similarly, the missing-energy range over which the data have been
integrated differs from case to case.  These differences have been
accounted for in the calculations used to interpret the data.  For
details on the analysis of \eep\ experiments and the extraction of
\rhopm\ from \eep\ cross-section data, the reader is referred to Ref.
\cite{her88}.

\section{Analysis of \onep\ knockout data}
\label{sec:onep}

\begin{figure} [t]
\epsfxsize=8cm
\centerline{\epsffile{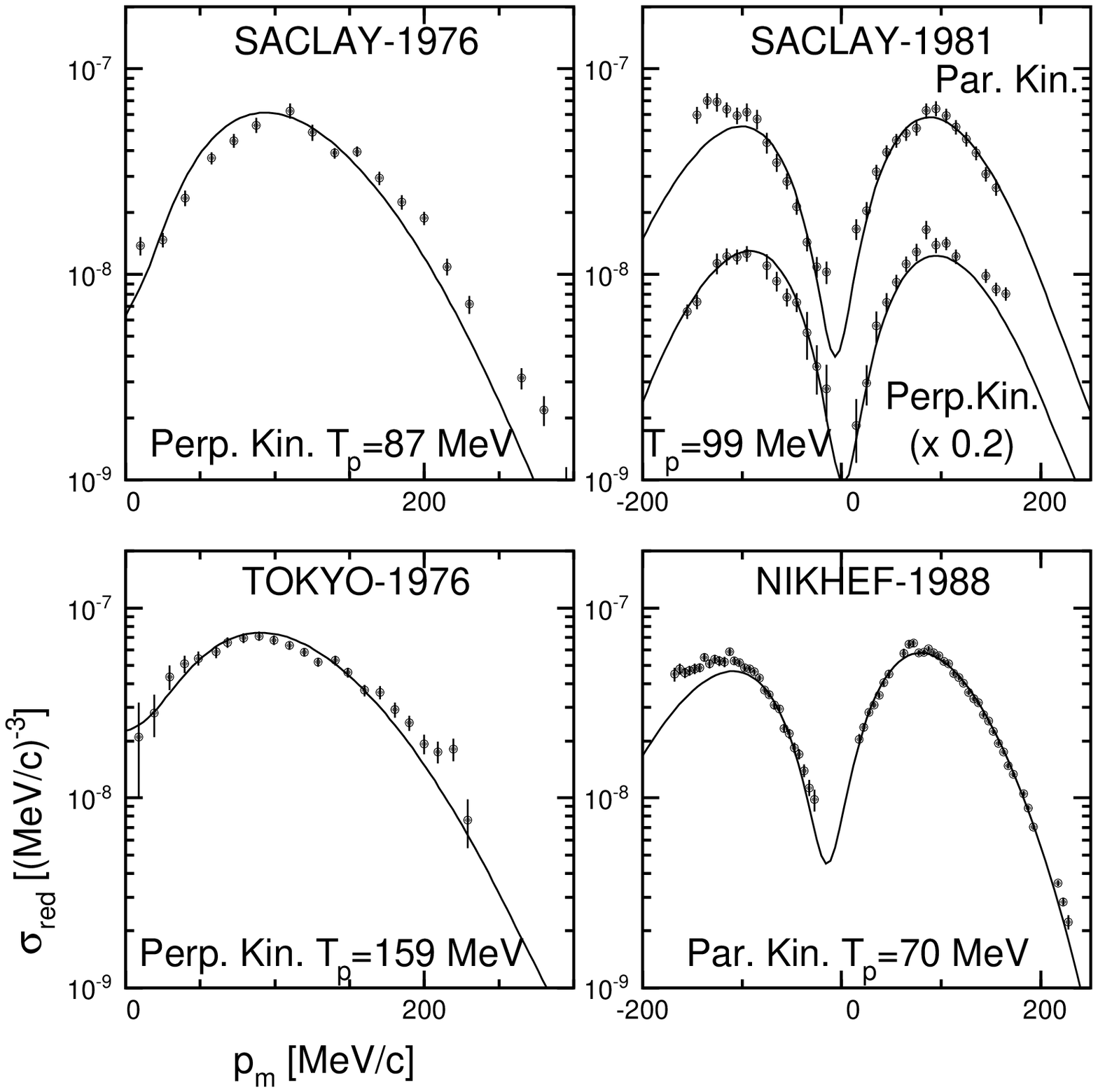}}
\caption{ Reduced cross sections for \onep -knockout from \carbon\ as
obtained with the reaction \carbon\eep.  The panels show data
collected in Tokyo \protect\cite{nak76}, Saclay
\protect\cite{mou76,ber82} and Amsterdam \protect\cite{ste88a} under
different kinematic conditions (see Table \ref{tab:kinem}).  The
data contain \onep\ transitions to the ground state and first and
second excited states in \boron.  The curves represent CDWIA
calculations summed over these transitions with spectroscopic factors
1.79, 0.22 and 0.19, respectively.  It is also noted that a radiative
correction has been applied to the Saclay data of Ref.
\protect\cite{ber82} as the published data were not corrected for
these effects.  }
\label{fig:all1p}
\end{figure}

In Fig.  \ref{fig:all1p} the \onep -knockout data from Refs.  
\cite{nak76,mou76,ber82,ste88a} are displayed and compared to Complete 
Distorted-Wave Impulse Approximation (CDWIA) calculations of the type 
described in Ref.  \cite{bof93}.  The input parameters of these 
calculations have been determined as follows.  The CDWIA calculations 
have been performed with a standard Woods-Saxon (WS) bound-state wave 
function and optical-potential parameters derived from elastic proton 
scattering off \carbon\ \cite{com80}.  The real part of the optical 
potential, which was thus interpolated from the tables of Ref.  
\cite{com80}, has been reduced by 5 MeV in order to account (partly) 
for channel-coupling effects.  (This procedure is verified in Ref.  
\cite{ste88a} by comparing to explicit coupled-channels calculations.)  
The calculated cross sections are divided by a kinematic factor and 
the electron-proton cross section \signr\ of McVoy and Van Hove 
\cite{mcv62}.  The use of \signr\ instead of \sigcc\ in the 
calculations is motivated by the fact that the nucleon-current 
operator in the CDWIA calculations is a non-relativistic expansion of 
the one that is used in \sigcc.  The division by \signr\ partly 
accounts for that difference.  (Note that in PWIA the correction is 
exact.)  For the kinematics of the experiments considered the ratio 
\signr\ $/$ \sigcc\ is between 0.95 and 0.98.  The spectroscopic 
factor \sonep\ and the radius $r_{0}$ of the WS well have been fitted 
to the data measured at NIKHEF for the \onepthree\ ground-state 
transition and \onepone\ and \onepthree\ transitions to the first two 
excited states, as these data have the smallest statistical and 
systematic uncertainties. The obtained fit values 
($S$, $r_{0}$ and $\chi^{2}/d.f.$) are 
(1.79$\pm$0.03, 3.12$\pm$0.05 fm, 165/34), 
(0.22$\pm$0.01, 3.94$\pm$0.05 fm, 52/37) and
(0.19$\pm$0.01, 3.34$\pm$0.06 fm, 47/37)
for the ground state, first and second excited state, respectively.  
These values are in agreement with those published before 
\cite{ste88a}.  Differences with previous values are due to minor 
changes in the CDWIA code, as described in Ref.  \cite{ire95}, and to 
the inclusion of an additional free parameter used in Ref.  
\cite{ire95}.  In more detailed analyses \cite{ste89,Giu88,Jes94} 
this parameter was shown to be unneeded to describe the data, 
whence we have omitted it in the present analysis.

The differences between the calculations and experimental data for 
negative \pmiss\ ($<$ -100 \mevc) in parallel kinematics (Saclay and 
NIKHEF data), are attributed to coupled-channels and charge-exchange 
effects, which are not included in the present analysis.  In Refs.  
\cite{ste89,Giu88,Jes94} it is shown that a good description of the 
momentum distribution at negative \pmiss can be obtained if these 
contributions, which are very small at positive \pmiss, are taken into 
account.  In order to avoid any bias of the presently deduced 
spectroscopic factors on the size of these contributions we have 
included the positive \pmiss\ data only in the fit to the NIKHEF data.  
Moreover, since the error bars of the negative \pmiss\ data are much 
larger than those of the positive \pmiss\ data the deduced 
spectroscopic factors are hardly affected by the omission of the 
negative \pmiss\ data in the fits.

Using the values of \sonep\ and $r_{0}$ as derived from the NIKHEF 
data, CDWIA calculations have been performed for the other data sets 
displayed in Fig.  \ref{fig:all1p}.  In each case the kinematics used 
as input for the calculations were adjusted to those used in the 
experiment, and the optical-model parameters were interpolated from 
the tables of Ref.  \cite{com80}.  For this purpose we used the proton 
laboratory scattering energy $T_{p}^{opt}$ as calculated via Eq. (4.3) 
of Ref.  \cite{bever} from the proton kinetic energy (\tp) employed in 
the experiment.  The aforementioned slight modification of the 
optical-model parameters was also applied.  This correction for 
channel-coupling effects presumably represents an overestimation as it 
was gauged at the lowest value of \tp, i.e.  70 MeV, where 
channel-coupling effects are largest.  However, since the effect of 
the channel-coupling correction on the deduced values of \sonep\ and 
$r_{0}$ for the dominant g.s.  transition is only 2\% or less at \tp =
70 MeV \cite{ste88a}, a more refined evaluation of channel-coupling 
effects at each value of \tp\ has not been carried out.  (Note that 
the channel-coupling effects are also small compared to the systematic 
uncertainty of the data, which ranges from 4\% to 15\%.)
It has to be realized that our procedure results in ${ absolute} $ 
calculations for all data sets, except the one obtained at NIKHEF that 
was used to fix the values of the spectroscopic factors and the radius 
of the bound-state wave functions.  

From Fig.  \ref{fig:all1p} it is concluded that the calculations give 
a fair simultaneous description of the data sets of Tokyo, Saclay and 
NIKHEF. The apparent discrepancy between the calculations and the 
Saclay data of 1976 at \pmiss $>$ 200 \mevc\ is probably related to an 
enhancement of the longitudinal-transverse interference structure 
function $W_{LT}$, which is absent in the data collected in parallel 
kinematics.  In the Saclay data of 1976, which were measured in 
\qom-constant (also called perpendicular) kinematics, an enhanced 
$W_{LT}$ term may show up at large \pmiss\ since its contribution to 
the cross section is proportional to sin$(\theta_{pq})$, where 
$\theta_{pq}$ is the angle between the three-momentum transfer and the 
outgoing proton momentum.  In Refs.  \cite{lap93,spal94,chin91} it has 
been shown for $^{16}$O and $^{40}$Ca that $W_{LT}$ is enhanced by up 
to a factor of two compared to standard CDWIA calculations.  Such an 
enhancement would only affect the \qom-constant data at high \pmiss\ 
and be stronger for small \tp.

\begin{figure} [t] \epsfxsize=8cm
\centerline{\epsffile{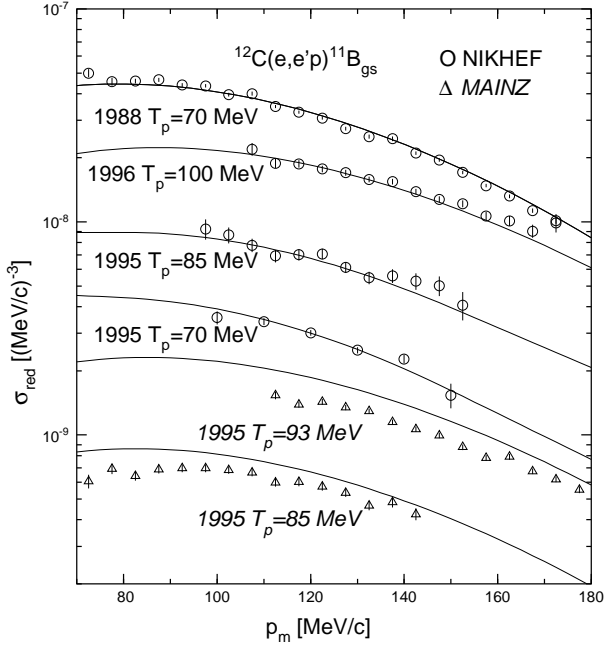}}
\caption{ Reduced cross sections for proton knockout from \carbon\
leading to the ground state \boron.  The shown data are those from an
early NIKHEF experiment \protect\cite{ste88a}, the present new NIKHEF
data and those of Mainz \protect\cite{blo95}.  The curves represent
CDWIA calculations for a ground-state spectroscopic factor of 1.79.
For clarity data and curves have been divided by consecutive factors
of 2, starting from the top.  }
\label{fig:plotgs}
\end{figure}

When we apply the absolute calculations, as described above, for the 
kinematics of the recently published \cite{blo95} Mainz experiment we 
find that their data lie about 20\% below the calculated reduced
cross sections (see Fig.  \ref{fig:plotgs}).  In order to resolve this 
discrepancy between the Mainz data and the other existing data, new 
measurements have been performed at the Amsterdam Pulse Stretcher 
(AmPS) facility \cite{wit93}.  The high duty-factor electron beams 
produced by AmPS enabled us to carry out high-statistics \carbon\eep\ 
measurements with hardly any contamination due to accidental 
coincidences in a short amount of time (less than 30 minutes) at an 
average beam intensity of 5 $\mu$A using a $102 \pm 1$ mg/cm$^{2}$ 
\carbon\ target.  The electron and the proton were detected and 
momentum-analyzed with two high-resolution magnetic spectrometers 
\cite{vri84}.  The kinematics of the measurements (summarized in Table 
\ref{tab:ampskin}) were chosen to be close to the kinematics of the 
existing \carbon\eep\ measurements described in Refs.  \cite{ste88a} 
and \cite{blo95}.  As the beam energy available differed somewhat from 
the value used in the two previous experiments, there is a small 
difference in the value of the virtual photon polarization parameter 
$\epsilon$.  However, as the ratio of the longitudinal and transverse 
response functions of the reaction \carbon\eep\ is known to be in 
agreement with the L$/$T ratio of the free electron-proton cross 
section \cite{ste91}, these differences are properly accounted for in 
the CDWIA calculations.

\begin{table}
\caption{Spectroscopic factors for the reaction \carbon\eep\ leading
to the ground state of \boron\ as determined from the present
experiments at NIKHEF and those of Mainz.}
\label{tab:ampskin}
\begin{center}
\begin{tabular}{|l|c|c|c|c|c|c|c}
\hline
& $E_0$ & $\Delta p_m$ & \tp & Kine-  & $S_{g.s.}$ & $\delta_{syst}$
\\
& MeV   & \mevc        & MeV & matics &            & \% \\
\hline
NIKHEF88 & 285-481 & -175,230 & 70 & par. &  $ 1.79 \pm 0.03$ & 4\\
NIKHEF95 & 378     & 100-150  & 70 & par. &  $ 1.79 \pm 0.04$ & 4\\
NIKHEF95 & 585     & 100-150  & 85 & par. &  $ 1.85 \pm 0.03$ & 4\\
NIKHEF96 & 611     & 100-150  & 100 & par. & $ 1.84 \pm 0.02$ & 4\\
Mainz95  & 855     & 70-140  & 85 & par. &  $ 1.50 \pm 0.02$ & 7\\
Mainz95  & 855     & 110-190 & 93 & par. &  $ 1.45 \pm 0.02$ & 7\\
\hline
\end{tabular}
\end{center}
\end{table}

The results of the new measurements are also shown in Fig.  
\ref{fig:plotgs}, where the data are compared to CDWIA calculations of 
the same type as described before, i.e.  the normalization 
($S_{g.s.}$=1.79) of the curves is derived from the data of Ref.  
\cite{ste88a}, while the optical-potential parameters and kinematics 
are properly derived from the experimental conditions.  Again a good 
description of the experimental data is found, thus confirming the 
normalization of the older experiments - from Refs.  
\cite{nak76,mou76,ber82,ste88a} - of Fig.  \ref{fig:all1p}.  If we fit 
the normalization of the curves to the experimental data we arrive at 
ground-state spectroscopic factors for each experiment as listed in 
Table \ref{tab:ampskin}.

Having established the proper normalization of most of the existing 
\carbon\eep\ data, we may now use these data as a collection of 
independent measurements of the nuclear overlap matrix element for the 
removal of \onep\ protons from \carbon\ leading to the ground state 
and low-lying excited states of \boron.  Hence, each of the data sets 
was used in order to determine a value of the spectroscopic factor 
\sonep\ for \onep -knockout from \carbon.  This has been done by 
fitting the data of each experiment with the corresponding CDWIA 
curves using \sonep\ as a free parameter.  The resulting values of 
\sonep\ are listed in Table \ref{tab:results}, and are seen to be in 
good agreement with each other.  As the individual values of \sonep\ 
have been derived from experiments that were carried out under widely 
different conditions, it is concluded that the treatment of the \eep\ 
reaction mechanism is well under control.  Further evidence supporting 
the validity of the CDWIA approach can be found in Ref.  \cite{ire94}, 
where it is shown that CDWIA calculations reproduce the nuclear 
transparency for protons at modest values of \QSQ\ (and thus of \tp), 
as measured at MIT/Bates \cite{gari92}.  Hence, by taking the weighed 
average of these independent values of \sonep\ (where the systematic 
uncertainties have been added quadratically to the statistical errors, 
see Table \ref{tab:results}) a good and reliable measure of the 
nuclear overlap matrix element is obtained, i.e.  \sonep = 2.23 $\pm$ 
0.07.

\begin{table}
\label{tab:results}
\caption{Experimental values of spectroscopic factors for \onep- and
\ones-knockout deduced for the various data sets from a fit with CDWIA
reduced cross sections.  The columns represent data set, \onep\
spectroscopic factor, \emiss-range for the deduced \ones\
spectroscopic factor, \ones\ spectroscopic factor and systematic error
$\delta_{syst}$ of the data set.  The listed uncertainties of the
spectroscopic factors do not include the contribution of
$\delta_{syst}$.}
\begin{center}
\begin{tabular}{|l|c|c|c|c|}\hline
data set&   \sonep & $\Delta E_m^{1s}$ & \sones & $\delta_{syst}$ \\
        &           &  MeV                  &        & \%       \\
\hline
Tokyo \cite{nak76} & 2.16 $\pm$ 0.10 & 21-30 & 0.08 $\pm$ 0.02 & 8 \\
Tokyo \cite{nak76} &                 & 30-42 & 0.66 $\pm$ 0.02     &
8   \\
Tokyo \cite{nak76} &                 & 42-54 & 0.36 $\pm$ 0.03     &
8   \\
Tokyo \cite{nak76} &                 & 54-66 & 0.09 $\pm$ 0.02     &
8   \\
Tokyo \cite{nak76} &                 & 21-66 & 1.19 $\pm$ 0.05     &
8   \\
Saclay \cite{mou76}& 2.19 $\pm$ 0.13 & 30-50 & 0.84 $\pm$ 0.02     &
15 \\
Saclay \cite{ber82}& 2.28 $\pm$ 0.07 &       &                   &  7
\\
Saclay \cite{ber82}& 2.31 $\pm$ 0.06 &       &                   &  7
\\
NIKHEF \cite{ste88a,ste88b}&2.20 $\pm$ 0.04 & 21-30 & 0.047 $\pm$ 0.002     &
4 \\
\hline
\end{tabular}
\end{center}
\end{table}

As compared to the independent-particle shell-model prediction (\sonep =4)
the value of \sonep\ (summed over the three \onep\ transitions) is 44
\% low, thus confirming the values earlier reported \cite{lap93,stew91},
albeit with higher precision.  Hence, the many-body interpretation of
the low spectroscopic factors found in \eep\ measurements at
\QSQ$<0.4$ \gevcsq\ need not be
revised.

\section{Analysis of \ones\ knockout data}
\label{sec:ones}

\begin{figure} [t]
\epsfxsize=8cm
\centerline{\epsffile{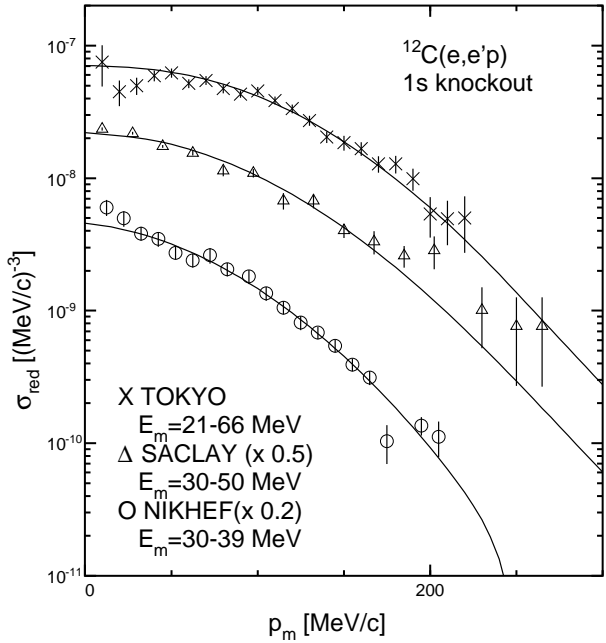}}
\caption{ Reduced cross sections for \ones -knockout from \carbon\ as
obtained with the reaction \carbon\eep.  The  data shown are those of
Tokyo \protect\cite{nak76}, Saclay \protect\cite{mou76} and NIKHEF
\protect\cite{ste88b} integrated over the indicated missing-energy
ranges (see Table \ref{tab:kinem}).  
The curves represent CDWIA calculations with a
spectroscopic factor fitted to the data.  }
\label{fig:s1s}
\end{figure}

Since the existing data for \ones\ knockout from \carbon\ cover 
different ranges in missing energy (see Table \ref{tab:kinem} and Fig.  
{\rm \ref{fig:s1s}}) and the experimental \ones\ missing-energy distribution 
extends over a range of about 25-80 MeV (see e.g.  Refs.  
\cite{nak76,mou76,lou86,ulm87,wein} a special procedure was followed 
to extract the \ones -strength.  Since the peak of the \ones\ 
missing-energy distribution is located at about 40 MeV we first fitted 
CDWIA calculations to the data of Saclay in the \emiss\ range 30-50 
MeV and those of Tokyo in the range 30-54 MeV. For these calculations 
we used a Woods-Saxon (WS) bound-state wave function with a binding 
energy of 40 MeV and fitted the radius of the WS well.  With the 
resulting geometry of the WS well ($r_{0}$=2.66 fm, $a_{0}$=0.65 fm) 
we calculated all other \ones\ reduced cross sections with wave 
functions that have a binding energy corresponding to the center of 
the missing-energy interval of the data under consideration.  Hence 
the depth of the well increases with increasing binding energy while 
simultaneously the rms radius of the wave function decreases.

\begin{figure} [t]
\epsfxsize=8cm
\centerline{\epsffile{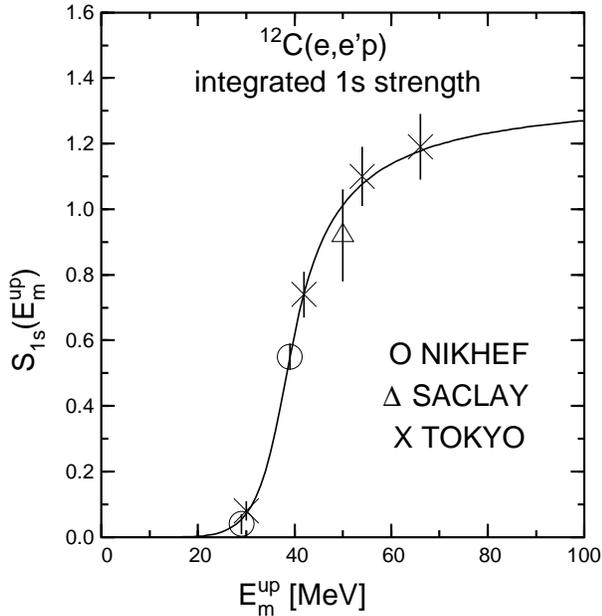}}
\caption{ Integrated \ones- knockout strength obtained with the
reaction \carbon\eep\ as a function of the upper integration limit in
missing energy.  The shown data are those of Tokyo
\protect\cite{nak76} (analyzed in four separate missing energy
intervals), Saclay \protect\cite{mou76} and NIKHEF
\protect\cite{ste88a}.  The curve represents a fit with an integrated
Lorentzian as described in the text.}
\label{figgpint}
\end{figure}

Next the normalization of these calculated \ones\ reduced cross 
sections was fitted to each data set to obtain the spectroscopic 
factor for \ones\ knockout in the particular interval (see Table 
\ref{tab:results} and Fig.  \ref{fig:s1s}).  Since the Tokyo data set 
in the interval \emiss =21-30 MeV contains both \ones\ and \onep\ 
strength a two-parameter fit was employed in this case.  From the 
obtained normalizations one can easily deduce the \ones\ strength 
\sones (\emup) integrated to an upper limit in missing energy denoted 
by \emup.  These values have been plotted in Fig.  \ref{figgpint} 
where the errors include statistical and systematic uncertainties (see 
Table \ref{tab:results}) added in quadrature.  The \ones\ strength at 
any value of \emup\ can now easily be deduced from a fit to the data 
with the expression :
\begin{eqnarray}
S_{1s}(E_{m}^{up}) &=& n_{1s} \int_{E_{F}}^{E_{m}^{up}} dE_{m}
\frac{\Gamma(E_{m})/2\pi}{(E_{m}-E_{1s})^{2}+\frac{1}{4}\Gamma^{2}(E_{m})},
\end{eqnarray}
where
\begin{eqnarray}
\Gamma(E_{m}) &=& \frac{a(E_{m}-E_{F})^{2}}{b+ (E_{m}-E_{F})^{2} }.
\label{eq:lorentzian}
\end{eqnarray}
In this approach the energy dependence of the spectral function is
modeled as a Lorentzian with an energy-dependent width
$\Gamma$(\emiss) that was calculated according to the formula given by
Brown and Rho \cite{BroR81} who use $a$=24 MeV and $b$=500 MeV$^{2}$.
In Eq. (1) the quantity $n_{1s}$ is the asymptotic (\emup $\rightarrow
\infty $) occupation for the \ones -shell, while $E_{1s}$ is the
centroid energy for the \ones -shell.  In the fit $b$, $n_{1s}$ and
$E_{1s}$ were treated as free parameters and found to be $b=590 \pm
250 $ MeV$^{2}$, $n_{1s}=1.32 \pm 0.08$ and $E_{1s}=39 \pm 1 $ MeV.
The deduced spreading width $\Gamma(E_{1s})=12 \pm 3$ MeV is in good
agreement with the broadening of the \ones\ missing-energy
distributions as shown in the Saclay \cite{mou76} and Tokyo
\cite{nak76} data.  The fitted curve is seen to describe the data
nicely.  For the analysis of the \ones\ SLAC data (\emiss =30-80 MeV)
we will employ the value \sones(80)-\sones(30)=$1.18 \pm 0.07$, where
all correlated errors in the fitted parameters have been included.

\section{Comparison with the SLAC data}
\label{sec:slacdata}

\subsection{Transparencies}

Using the precisely determined values of \sonep\ and \sones\ we have
also reconsidered the interpretation of the \carbon\eep\ experiment
performed at SLAC \cite{mak94} at a somewhat higher value of \QSQ =
1.1 \gevcsq.  In this experiment the nuclear transparency for protons
was measured with the aim of searching for color-transparency effects
\cite{MulB}, i.e.  the predicted increase of the nuclear transparency
due to the proposed reduced interaction probability of small color
neutral objects with the surrounding medium (see Ref.  \cite{CT97} for
a recent review).

The experimental nuclear transparency $T_{\alpha}$ ($\alpha=1s,1p$) is 
determined by fitting a PWIA curve to the data using its normalization 
as a free parameter.  However, as the magnitude of the PWIA curve 
scales with both $S_{\alpha}$ and $T_{\alpha}$, any uncertainty in 
$S_{\alpha}$ is immediately reflected in the derived value of 
$T_{\alpha}$.  In Refs.  \cite{gees,mak94,neith}, {\it theoretical} 
estimates for \sonep\ and \sones\ were used, creating a theoretical 
bias in the derived values of \trans.  With the presently available 
precise values of \sonep\ and \sones\ in hand, it is now possible to 
derive a value for the nuclear transparency that is based on {\it 
experimental} results for the spectroscopic factors.  It is noted that 
this procedure relies on the assumption that the reduction of 
spectroscopic strength (to about 60\% of the IPSM value), which we
derived from the low \QSQ\ measurements, is the same at \QSQ = 1.1 
\gevcsq.  This implies that the increase of \QSQ\ does not affect the 
amount of strength residing in the acceptance of the experiment 
(\emiss $<$ 80 MeV).  Future experiments with a larger acceptance (and 
very good signal-to-noise ratios!)  can in principle study the 
validity of this assumption by searching for strength at high missing 
energies.

\begin{figure} [t] \epsfxsize=8cm
\centerline{\epsffile{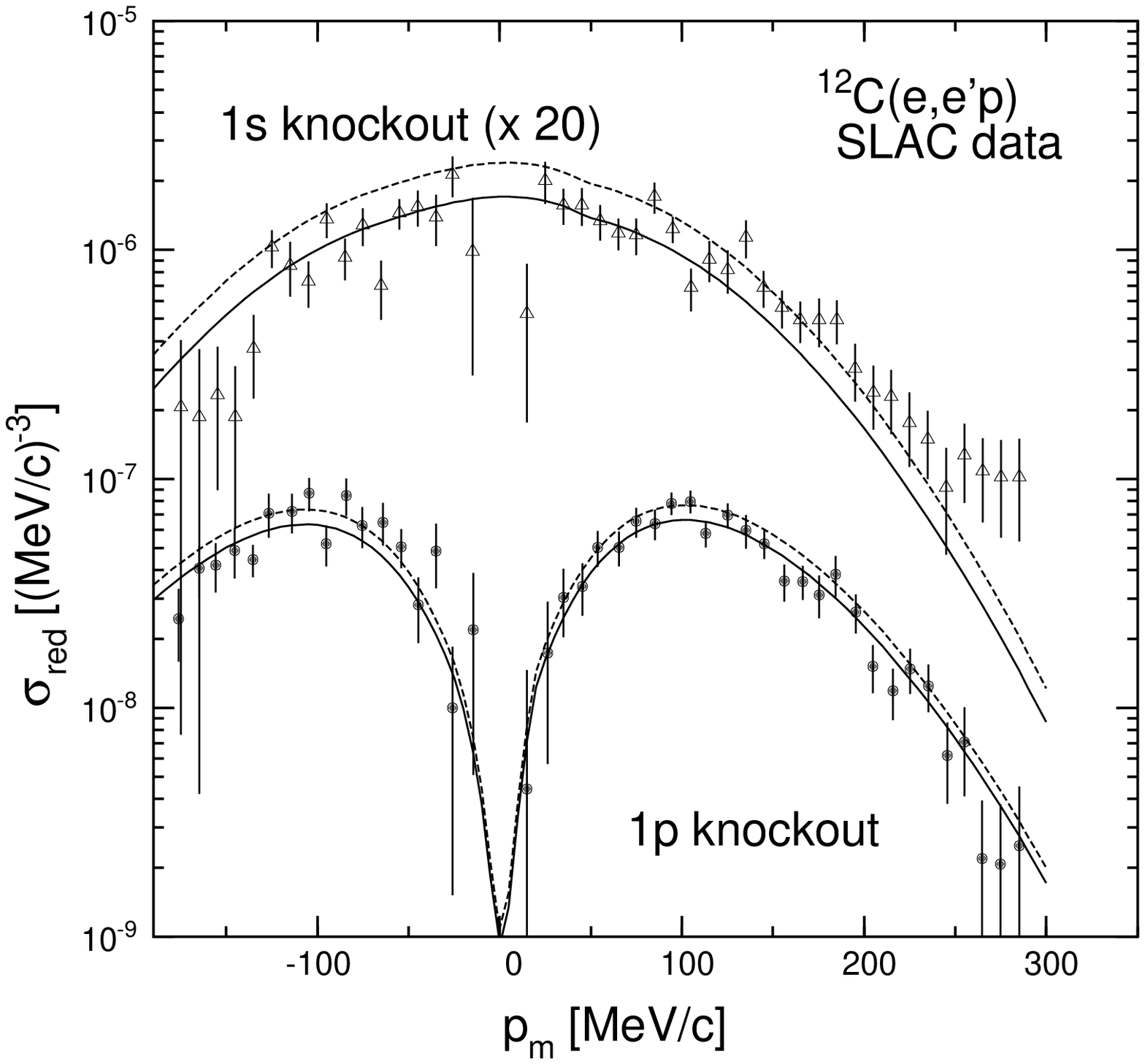}}
\caption{ Reduced cross section for \onep\ and \ones\ proton knockout 
in the reaction \carbon\eep\ as obtained in a recent SLAC experiment 
at \QSQ = 1.1 \gevcsq\ (from Ref.  \protect\cite{neith}).  The 
kinematics are given in Table \protect\ref{tab:kinem}.  The dashed 
curves (which assume a 100\% transparent nuclear medium (\trans=1)) 
represent PWIA calculations normalized with the spectroscopic factors 
\sonep\ = 2.23 and \sones\ = 1.18 derived from the world's 
\carbon\eep\ data displayed in Figs.  \protect\ref{fig:all1p} and 
\protect\ref{fig:s1s}.  For the solid curves the transparency has 
been fitted to the data.  }
\label{fig:plotslac}
\end{figure}

We have obtained the SLAC \carbon\eep\ data for \onep- and 
\ones-knockout from Ref.  \cite{neith}, and applied radiative 
corrections to these data.  The size of these corrections coincides to 
within 2.5\% with those calculated by the authors of Ref.
\cite{neith}.  The SLAC \onep - and \ones -knockout data for \QSQ = 
1.1 \gevcsq\ are displayed in Fig.  \ref{fig:plotslac}, where they are 
compared to a plane-wave impulse approximation calculation (PWIA) 
based on the BSWF-parameters and spectroscopic factors derived from 
the low \QSQ\ data that were described in sections \ref{sec:onep} and 
\ref{sec:ones}.  Hence, final-state interaction effects are neglected.  
The PWIA curves are in reasonable agreement with the data, immediately 
suggesting a relatively large value of \trans.  Subsequently the data 
were fitted with the expression

\begin{equation}
\sigma^{exp}_{red}(p_m) = T_{\alpha} \cdot \sigma^{PWIA}_{red}(p_m)
\end{equation}
where $T_{\alpha}$ is treated as a free parameter.  This procedure
yields \transp\ = 0.86 $\pm$ 0.05 and \transs\ = 0.71 $\pm$ 0.06.  For
the data in the region \emiss =30-80 MeV we employed a fit using both
a \ones\ and a (small) \onep\ component.  The presence of the latter
is due to the fact that the SLAC data are not radiatively unfolded and
hence the radiative tail of the \onep\ distribution (which has an
exactly calculable magnitude) is also included in this energy region.

Combining the two results for the transparency of \onep -shell and
\ones -shell protons, we have evaluated the average transparency of
nucleons removed from \carbon\ according to
\begin{equation}
\label{eq:ttotal}
{ T_{^{12}C} } = { { S_{1p} T_{1p} + S_{1s} T_{1s} }
\over { S_{1p} + S_{1s} } }
\end{equation}
yielding $T_{^{12}C}$ = 0.81$\pm$0.04, which is considerably larger
than the value 0.65$\pm$0.05 quoted in Ref.  \cite{mak94}.

The origin of this difference is due to the way the authors of Ref.  
\cite{mak94} analyze their data.  First, they determine an overall 
proton transparency for the data integrated up to \emiss=100 MeV, 
whereas we deduce separate (and significantly different) 
transparencies for the \onep- and \ones-shell, and then obtain the 
weighed average.  Secondly, they use an (overall) {\it theoretical} 
'correlation correction' of 0.901$\pm$0.024 to normalize their PWIA 
calculation, whereas we use {\it experimentally} determined 
spectroscopic factors to separately normalize the \onep\ and \ones\ 
momentum distributions by 0.56$\pm$0.02 and 0.59$\pm$0.04, 
respectively.  Thirdly, the authors of Ref.  \cite{mak94} use 
bound-state wave functions generated in a Woods-Saxon potential 
derived from an early analysis of the Saclay data \cite{mou76}, which 
did not include non-locality corrections, Coulomb distortion and 
off-shell effects and used an optical potential without a spin-orbit 
term.  In our treatment the bound-state wave functions are based on an 
analysis of the world's data set for the reaction \carbon\eep, which 
accounts for all these effects and moreover uses an optical potential 
that describes proton scattering in the full employed proton energy 
range and includes corrections for coupled-channels effects.  As a 
result the bound-state wave functions are different (the rms radius 
for the \onep (\ones) wave functions differs by +7 (-2) \%).  These
three reasons explain why the value $T_{^{12}C}$=0.81$\pm$0.04 that we 
deduce in the present analysis is significantly larger than the value 
0.65$\pm$0.05 obtained in Ref.  \cite{mak94}.  As our value for 
$T_{^{12}C}$ has been obtained in the \emiss\ range up to 80 MeV, 
whereas the overall SLAC value was obtained from data integrated up to 
100 MeV, one may wonder whether this difference could explain the 
difference in obtained transparency.  Inspection of the measured SLAC 
missing-energy distribution \cite{mak94,neith} shows that in the range 
\emiss =80-100 MeV it closely follows their simulated theoretical 
curve and hence their deduced transparency value is not sensitive 
to the choice of the upper \emiss\ integration limit.

\begin{figure} [t]
\epsfxsize=8cm
\centerline{\epsffile{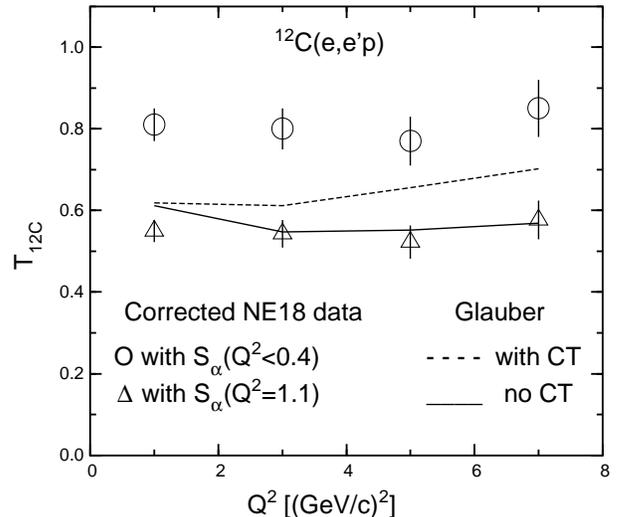}}
\caption{Transparency for the SLAC NE18 data on \carbon\ as a 
function of \QSQ. The circles (triangles) represent the values 
obtained when the spectroscopic factors determined at \QSQ$<$0.4 
\gevcsq\ (\QSQ=1.1 \gevcsq) are used.  The solid (dashed) curves 
represent Glauber calculations of Zhalov \etal\ \protect\cite{zhalov} 
without (with) Color Transparency.  }
\label{fig:newslac}
\end{figure}

The SLAC data also include \carbon\eep\ measurements at \QSQ\ values 
of 3, 5 and 7 \gevcsq.  In order to derive proper values of the 
nuclear transparency in this \QSQ\ range, the spectroscopic factors 
for \onep\ and \ones\ knockout quoted above should be used again.  
Rather than carrying out the same analysis for the higher \QSQ\ data 
again, we have used the ratio of the presently obtained value for 
$T_{^{12}C}$ and the published value of $T_{^{12}C}$ as a correction 
factor.  This simplified procedure is motivated by the fact that the 
difference between the published data and the present analysis is 
largely due to use of experimentally constrained spectroscopic 
factors.  We have thus applied the factor $F_{corr} = T_{^{12}C}^{new} 
/ T_{^{12}C}^{NE18}$ (as derived from the \QSQ\ = 1.1 \gevcsq\ data) 
to the other NE18 data, the result of which is displayed in Fig.  
\ref{fig:newslac} by the circle symbols.  An average nuclear 
transparency of about 0.8 is found.

The corrected NE18 data are compared to two Glauber calculations for 
the transparency.  The solid curve is a standard Glauber calculation, 
while the dashed curve includes Color Transparency effects.  Both 
calculations assume that the spectral strength has reached its 
asymptotic value, i.e.  no corrections are made for possible knockout 
strength outside the range of the experiment. We observe that the 
corrected NE18 transparency data are well above the calculations, 
especially at the lowest \QSQ\ values.  As Color Transparency is not 
expected to have a significant influence on the data below $Q^2 
\approx$ 4 \gevcsq\ (see also Refs.  \cite{nikol,benhar}), the 
discrepancy at \QSQ=1.1 \gevcsq\ is particularly disturbing.

It is difficult to identify a possible origin for the 4.2 $\sigma$
deviation between the corrected data and the Glauber calculation at
\QSQ\ = 1.1 \gevcsq, because of the following reasons: (i) the recent
TJNAF data for the nuclear transparency \cite{gees} confirmed the NE18
data (using the same value for the 'correlation correction'), making
it unlikely that the effect is due to an experimental error; (ii) the
optical model calculations for the low-energy \carbon\eep\ data give
consistent results for different kinematics, different nuclei
\cite{lap93,stew91}, and are even able to reproduce the measured
nuclear transparency in the very low \QSQ\ domain \cite{ire94}; (iii)
it is hard to believe that the Glauber calculations are incorrect as
they are able to reproduce elastic and inelastic proton scattering
data in the relevant energy domain (few GeV) \cite{glauber}, and
different authors are able to reproduce the theoretical calculations
shown in Fig. \ref{fig:newslac} \cite{benhar,nikol,zhalov}.

\begin{figure} [t] \epsfxsize=8cm
\centerline{\epsffile{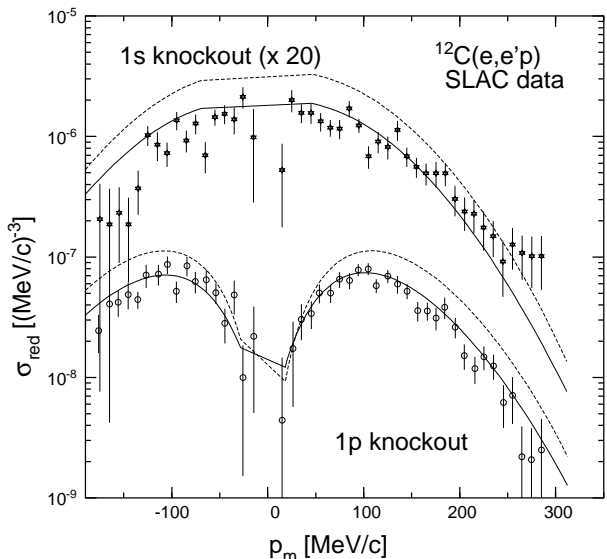}}
\caption{ Reduced cross section for \onep\ and \ones\ proton knockout 
in the reaction \carbon\eep\ as obtained in a recent SLAC experiment 
at \QSQ = 1.1 \gevcsq\ (from Ref.  \protect\cite{neith}).  The 
kinematics are given in Table \protect\ref{tab:kinem}.  The curves 
represent momentum distributions calculated in PWIA (dashed) and in 
the Glauber approximation (solid).  For all curves the spectroscopic 
factors $S=2j+1$ were employed.  }
\label{fig:plotzhalov}
\end{figure}

\subsection{Spectroscopic factors}

If one trusts the Glauber approach as a reliable calculation of the 
final-state interaction at high \QSQ, the SLAC data may be used 
instead to extract spectroscopic factors at these momentum transfers.  
For that purpose we carried out Glauber calculations for the SLAC 
momentum distributions, which are compared to the data in Fig.  
\ref{fig:plotzhalov}.  Apart from the Glauber calculation itself 
(solid curve) also a PWIA curve (dashed line) is shown.  For both 
curves the spectroscopic factors were set equal to $S=2j+1$.  One 
immediately derives from the ratio between Glauber and PWIA curves 
that at this energy the absorption factor due to the final-state 
interaction for \onep\ (\ones) knockout is 0.6-0.7 (0.5-0.6), where 
the range indicates the dependence on \pmiss.  At first glance the 
Glauber curves with $S=2j+1$ seem to describe the data rather well, 
but if one fits the data with the \pmiss\ dependence of the Glauber 
curves one arrives at spectroscopic factors \sonep=3.56$\pm$0.12 and 
\sones= 1.50$\pm$0.08.  These values are appreciably larger than the 
ones determined from the analysis of the world's low \QSQ\ data as 
presented in sections \ref{sec:onep} and \ref{sec:ones}.

Obviously, when we apply these spectroscopic factors in the 
calculation of the PWIA momentum distribution (see Eq.  (3)) in order 
to determine the transparency, we arrive at the much lower 
transparency values indicated in Fig.  \ref{fig:newslac} by the 
triangles.  These values are close to the original NE18 values since 
the total spectroscopic strength, \sonep + \sones=5.06$\pm$0.14, 
determined here from the Glauber fits, is close to the theoretical 
value $6 \times 0.901 \pm 0.024 =5.41 \pm 0.14$ employed in the 
original NE18 analysis.

\section{\QSQ\ dependence of the deduced spectroscopic strength}
\label{sec:qdependence}

\begin{figure} [t] \epsfxsize=7cm
\centerline{\epsffile{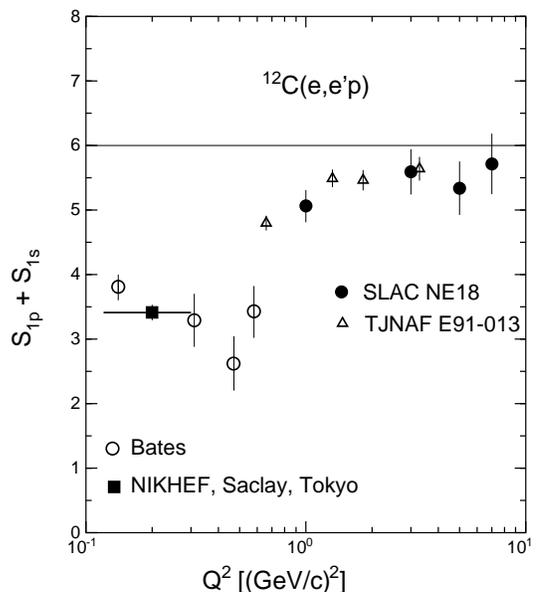}}
\caption{\QSQ\ dependence of the summed spectroscopic strength 
$S_{1p}+S_{1s}$ for \onep\ and \ones\ proton knockout in the reaction 
\carbon\eep\ up to \emiss = 80 MeV. The square indicates the result 
from the combined analysis of NIKHEF, Saclay and Tokyo data (see 
sections \protect\ref{sec:onep} and \protect\ref{sec:ones}), where the 
horizontal bar denotes the \QSQ\ range of these data.  Other symbols, 
as indicated, represent the results from experiments at Bates 
\protect\cite{ulm87,wein}, SLAC \protect\cite{mak94} and TJNAF 
\protect\cite{gees}.  }
\label{fig:plotspfac}
\end{figure}

The apparent discrepancy between the analysis of \carbon\eep\ data at 
low and at high \QSQ\ is illustrated in Fig.  \ref{fig:plotspfac}.  
Here, we plot the summed spectroscopic factors \sonep +\sones\ for 
\onep\ and \ones\ knockout as a function of \QSQ\ in the range between 
0.1 and 10 \gevcsq.  At low \QSQ ($<$ 0.6 \gevcsq) the results of the 
combined analysis of the NIKHEF, Saclay and Tokyo data (see sections 
\ref{sec:onep} and \ref{sec:ones}) are shown and those of two 
experiments performed at Bates \cite{ulm87,wein}, which covered a 
small \pmiss\ acceptance and were therefore not included in the analysis 
of sections \ref{sec:onep} and \ref{sec:ones}.  All low \QSQ\ results, 
which were obtained with an optical-model treatment of the final-state 
interaction, are mutually consistent and lead to a total strength 
\sonep +\sones =3.45$\pm$0.13. At higher \QSQ\ we plot the data of 
the SLAC experiment \cite{mak94}, as discussed in section 
\ref{sec:slacdata}, and those of a recently published TJNAF experiment 
\cite{gees}.  Here, the spectroscopic factors were deduced from a 
comparison of experimental cross sections with calculations employing 
a Glauber approach for the final-state interaction.  These data 
exhibit a modest \QSQ\ dependence, which is already interesting in 
itself, and moreover, they do not seem to join smoothly to the low 
\QSQ\ data.

In conventional nuclear-structure models the spectroscopic strength 
should be independent of \QSQ. Hence, the question arises what the 
origin of the observed discontinuity near \QSQ =0.6 \gevcsq\ can be.  
The two main differences in the analysis of the low and the high \QSQ\ 
data are a different treatment of the final-state interaction and the 
use of a different current operator.  Kelly \cite{kelly96} has 
calculated the final-state interaction in the \QSQ\ range 0.2-1.2 
\gevcsq\ using an optical model with the EEI interaction, which was 
compared \cite{gees} to the results of a calculation involving the 
Glauber approach.  Differences between the two approaches of up to 
10\% are found, but these are not sufficient to explain the observed
discontinuity.

The current operator used in the analysis of the low \QSQ\ data is a 
non-relativistic one, whereas in the Glauber calculations performed 
for the analysis of the high \QSQ\ data a relativistic current 
operator is employed.  Earlier comparisons \cite{Jin93,udias} of 
relativistic versus non-relativistic analyses \eep\ data at low \QSQ\ 
have shown that differences in the extracted spectroscopic strength of 
up to 15\% occur, again not enough to explain the observed
discrepancy.  Clearly, a consistent analysis of all data between 0.1 
and 10 \gevcsq\ could improve insight into this matter.  Such an 
analysis is beyond the scope of the present paper.

Finally, it should be noted that in all analyses only one-body 
operators are included in the current operator.  Since two-body 
currents (meson exchange, intermediate delta excitation) markedly 
differ in their \QSQ\ dependence from the one-body current, these may 
be at the origin of the observed \QSQ\ dependence of the the extracted 
strength.  In a recent L/T separation of \carbon\eep\ data carried out 
at \QSQ = 0.6 and 1.8 \gevcsq\ at TJNAF \cite{dutta00} such a \QSQ\ 
dependence of the transverse response, which receives contributions 
from the two-body currents, has been observed.  Further calculations, 
involving one-body and two-body currents in the operator, are needed 
to quantify this contribution.  It should be noted, though, that the 
contributions due to meson-exchange currents and intermediate delta 
excitation are expected to become less important with increasing \QSQ, 
while at low \QSQ\ they have been estimated to be small \cite{mecref}.

\section{Discussion}
\label{sec:discussion}

The analysis of the various data sets presented in sections 
\ref{sec:onep}-\ref{sec:slacdata} has revealed an unexpected 
\QSQ\ dependence of the spectroscopic factors deduced from 
\carbon\eep\ experiments in the quasi-elastic domain.  In the previous 
section it has been argued that the observed \QSQ\ dependence 
(illustrated in Fig.  \ref{fig:plotspfac}) could be caused by changes 
in the mechanism of the \eep\ reaction with \QSQ. However, at this 
point it remains unclear whether such changes are large enough to 
explain the data, since all the effects discussed in section 
\ref{sec:qdependence} are constrained by other experimental data.  
Hence, it is worthwhile to consider other possible explanations of the 
remarkable \QSQ\ dependence of the data as well.

For instance, it could be speculated that spectroscopic factors have 
an intrinsic \QSQ\ dependence.  While such an ansatz is in conflict 
with conventional models of nuclear structure, other many-body systems 
are known to have a scale-dependent renormalization.  As an 
illustration we mention the quasi-particle description of many-body 
fermion systems in condensed matter physics, and the QCD description 
of the quark-gluon structure of the nucleon (see also Refs.  
\cite{LL,migdal,Ma}).  In fact, both in condensed matter physics and 
in nuclear physics a description of a many-body system in terms of 
quasi-particles interacting through an effective potential is quite 
successful \cite{LL,Brown}.  In such a description the 
correlations between particles are included by using effective 
potentials.  The size of these effective potentials changes if the 
resolution by which the system is probed increases, as some of the 
correlations are resolved.
  
In order to visualize how a possible \QSQ\ dependence of spectroscopic 
factors may come about we consider the relevant energy 
and time scales that are involved in quasi-elastic electron 
scattering.  The electron-nucleon interaction itself can be 
characterized by a time scale $\tau_{int} \approx \hbar / \omega$, 
where $\omega$ represents the energy transfer to the struck nucleon.  
The time scale that characterizes the binding of the nucleon with the 
nuclear mean field $U$ is given by $\tau_{bind} \approx \hbar / U$.  
If $\tau_{int}$ and $\tau_{bind}$ are of similar size the effects of 
nuclear binding (i.e., long-range correlations) will be important.  
This situation occurs at low values of \QSQ, i.e., those corresponding 
to the kinematics used in the Bates, NIKHEF, Saclay and Tokyo 
experiments, with $\tau_{int} \approx$ 2 fm/$c$ and $\tau_{bind} 
\approx$ 4 fm/$c$.  On the other hand, under the conditions used in the 
SLAC experiment (at \QSQ = 1 \gevcsq), $\tau_{int} \approx$ 0.2 
fm/$c$, while $\tau_{bind}$ remains unchanged.  Hence, the effect of 
long-range correlations has a tendency to disappear at high \QSQ, 
resulting in a rise of the spectroscopic factors with momentum transfer.

Although the argument given above explains the qualitative features of 
the observed \QSQ\ dependence of the spectroscopic factors, it is too 
early to draw definite conclusions.  A more quantitative evaluation of 
both the reaction mechanism effects and the proposed renormalizability 
of spectroscopic factors is needed for a full development of this 
subject.

\section{Conclusion}
\label{sec:summary}

A detailed analysis of existing \carbon\eep\ data has shown the mutual 
consistency of existing data sets - with the exception of recent data 
from Ref.  \cite{blo95}.  \footnote{In a private communication with 
representatives from the Mainz experiment, it has become clear that 
the deviation between the Mainz data and the data from other 
laboratories is presumably due to a complicated dead time effect.} New 
experimental data obtained at the high-duty factor AmPS facility 
confirm the normalization of the older data sets.  From all data 
available precise values for the spectroscopic factors for \onep\ and 
\ones\ proton knockout from \carbon\ have been derived, which were 
used to re-evaluate the nuclear transparency from \carbon\eep\ data 
measured at high \QSQ. The deduced nuclear transparency is 
considerably closer to unity than previously reported.  If we assume 
instead that Glauber calculations give an adequate description of the 
final-state interaction effects at high \QSQ, the same data can be 
used to derive independent values of the spectroscopic factor at high 
\QSQ. In such an approach the spectroscopic factors for proton 
knockout from \carbon\ show an unexpected \QSQ\ dependence.  We have 
discussed several possible explanations for this unexpected 
observation.  As there is no treatment of the \eep\ reaction mechanism 
available that can be consistently applied from \QSQ = 0.1 to 10 
\gevcsq, it cannot be excluded that the \QSQ\ dependence of the 
spectroscopic factors is an artifact of the reaction mechanism 
description.  On the other hand it can also be speculated that the 
spectroscopic factors have an intrinsic \QSQ\ dependence, due to the 
possibly reduced influence of long-range correlations at high \QSQ. 
Further calculations are called for to resolve this issue.

\section*{Acknowledgments}

The authors like to thank M.F. van Batenburg and R. Medaglia for their 
contributions to the analysis of the AmPS \carbon\eep\ data.  They 
also acknowledge fruitful communications with Dr.  T.G.  O'Neill, Prof.  
N.C.R. Makins, and Prof.  R.G. Milner on the interpretation of the 
NE18 data.  Two of us (L.L., M.S.) want to thank the Institute for Nuclear 
Theory at the University of Washington for its hospitality and the 
Department of Energy for partial support during completion of this 
work.  This work is part of the research program of the Stichting voor 
Fundamenteel Onderzoek der Materie (FOM), which is financially 
supported by the Nederlandse Organisatie voor Wetenschappelijk 
Onderzoek (NWO).  M.S. would like to thank DESY for the hospitality 
during the time this work was done.  Research of M.S. was supported in 
part by the U.S. Department of Energy, research of L.F. was supported by 
the Israely Academy of Science.

\end{document}